\documentclass[amsmath,amssymb,reprint,aps,prl,superscriptaddress,showpacs]{revtex4-1}

\usepackage{graphicx} 
\usepackage{dcolumn}  
\usepackage{bm}       

\begin{document}

\title{CaIrO$_3$ post-perovskite, a $j\!=\!1/2$ quasi-one-dimensional antiferromagnet}

\author{Nikolay A.~Bogdanov}
\affiliation{Institute for Theoretical Solid State Physics, IFW Dresden, Helmholtzstr. 20, 01069 Dresden, Germany}

\author{Vamshi M.~Katukuri}
\affiliation{Institute for Theoretical Solid State Physics, IFW Dresden, Helmholtzstr. 20, 01069 Dresden, Germany}

\author{Hermann Stoll}
\affiliation{Institute for Theoretical Chemistry, Universit\"{a}t Stuttgart, Pfaffenwaldring 55, 70550 Stuttgart, Germany}

\author{Jeroen van den Brink}
\affiliation{Institute for Theoretical Solid State Physics, IFW Dresden, Helmholtzstr. 20, 01069 Dresden, Germany}

\author{Liviu Hozoi}
\affiliation{Institute for Theoretical Solid State Physics, IFW Dresden, Helmholtzstr. 20, 01069 Dresden, Germany}

\date{\today}


\begin{abstract}

The $5d^5$ iridate CaIrO$_3$ is isostructural with the post-perovskite phase of MgSiO$_3$,
recently shown to occur under extreme pressure in the lower Earth's mantle.
It therefore serves as an analogue of post-perovskite MgSiO$_3$ for a wide variety
of measurements at ambient conditions or achievable with conventional multianvile pressure
modules.
By multireference configuration-interaction calculations we here provide
essential information on the chemical bonding and magnetic interactions in CaIrO$_3$.
We predict a large antiferromagnetic superexchange 
of 120 meV along the $c$ axis,
the same size with the interactions in the 
cuprate superconductors,
and ferromagnetic couplings smaller by an order of magnitude along $a$.
CaIrO$_3$ can thus be regarded as a $j\!=\!1/2$ quasi-one-dimensional antiferromagnet.
While this qualitatively agrees with the stripy magnetic structure proposed by resonant
x-ray diffraction, the detailed microscopic picture emerging from our study,
in particular, the highly uneven admixture of $t_{2g}$ components,
provides a clear prediction for resonant inelastic x-ray scattering experiments.
\end{abstract}

\maketitle


Iridium oxide compounds are at the heart of intensive experimental and theoretical investigations
in solid state physics.
The few different structural varieties displaying Ir ions in octahedral coordination and
tetravalent $5d^5$ valence states, in particular, have recently become a fertile ground for
studies of new physics driven by the interplay between strong spin-orbit interactions and
electron correlations \cite{214Ir_kim_2009}.
The spin-orbit couplings (SOC's) are sufficiently large to split the Ir $t_{2g}^5$
manifold into well separated $j\!=\!3/2$ and $j\!=\!1/2$ states.
The former are fully occupied while the latter are only half filled.
It appears that the widths of the $j\!=\!1/2$ bands, $w\!\sim\!Nt$, where $t$ is an effective
Ir-Ir hopping integral and $N$ is the coordination number, are comparable to the strength
of the onsite Coulomb repulsion $U$.
For smaller coordination numbers, such as $N\!=\!3$ in the layered honeycomb systems 
Li$_2$IrO$_3$ and Na$_2$IrO$_3$ and $N\!=\!4$ for square-lattice Sr$_2$IrO$_4$ and Ba$_2$IrO$_4$
and the post-perovskite CaIrO$_3$, the undoped compounds are insulating at all temperatures and a 
$j\!=\!1/2$ Mott-Hubbard type picture seems to be appropriate
\cite{214Ir_kim_2009,213Ir_singh_2012,CaIrO3_ohgushi_2006}.
For larger coordination numbers like $N\!=\!6$ in the pyrochlore structure, the $t_{2g}^5$
Ir oxides are metallic at high T's and display transitions to antiferromagnetic (AF)
insulating states on cooling \cite{227Ir_matsuhira_2011}.
Here, a Slater type picture in which the gap opening is intimately related to the onset of
AF order might be more appropriate, as proposed for example for Os oxide perovskite systems with
$N\!=\!6$ \cite{NaOsO3_calder_2012}.
Nevertheless, also more exotic ground states such as a Mott topological insulator \cite{227Ir_pesin_2010} or
a Weyl semimetal \cite{227Ir_savrasov_2011} have been proposed for the pyrochlore iridates.

\begin{figure}[!b]
\includegraphics[width=0.90\columnwidth]{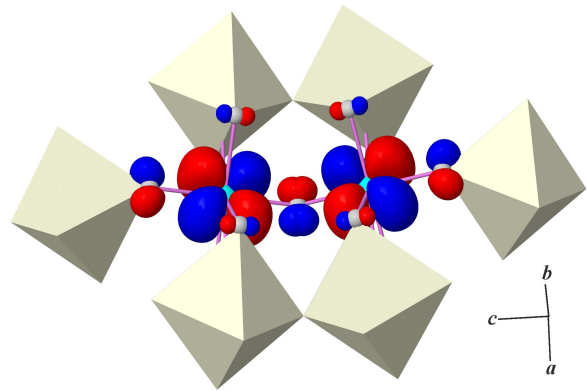}
\caption{
Local coordination and the main, $\pi$-type $5d_{xz}$-$2p_x$-$5d_{xz}$ AF superexchange path
for corner-sharing IrO$_6$ octahedra along the $c$ axis in CaIrO$_3$, see text.
Nearest-neighbor O ions around the two Ir sites in the center of the figure are shown as small light-grey
spheres.
Ca ions above and below the IrO$_3$ layer are not displayed.
The O $2p$ tails of the active Ir $d_{xz}$ orbitals are overlapping at the bridging O site.
}
\label{fig01}
\end{figure}

Besides the SOC's and Coulomb interactions, one important additional energy scale is the magnitude
of the splittings within the Ir $t_{2g}$ shell.
Such splittings may arise from distortions of the O octahedron around the Ir site and also
from the anisotropy of the farther surroundings.
In this respect, highly anisotropic systems such as the post-perovskite CaIrO$_3$ provide
an ideal playground for new insights into the interplay between SOC's and lattice distortions.
In the layered post-perovskite structure, IrO$_6$ octahedra in the $ac$ layers share corners
along the crystallographic $c$ axis, see Fig.~\ref{fig01}, and have common edges along $a$ \cite{CaIrO3_rodi_1965}.
The Ir-O bonds along the corner-sharing chains are substantially shorter than for the edge-sharing
links, 1.94 vs.~2.07  \AA, which suggests a stabilization of the $xy$-like orbital with respect
to the $xz$ and $yz$ terms.
Additionally, small splittings between the $xz$ and $yz$ components may in principle arise due
to the $a$-$b$ crystal anisotropy.

The precise nature of the relativistic $5d^5$ ground-state wave function obviously depends on
the size of the Ir $t_{2g}$ splittings.
Further, the on-site $5d^5$ electron configuration and the kind of Ir-O-Ir path along a given
direction determine the sign and magnitude of the magnetic exchange couplings.
{\it Ab initio} calculations where all these effects and interactions are treated on equal footing
are therefore highly desirable.  
For this, we focus on the post-perovskite phase of CaIrO$_3$, which has received significant attention
from geologists in recent years because the same crystalline structure is adopted by MgSiO$_3$ in 
the lower Earth's mantle \cite{pPv_mgsio3_murakami04,pPv_mgsio3_oganov04}.
In contrast to MgSiO$_3$, the post-perovskite phase of CaIrO$_3$ can be synthesized even at ambient
pressure.
Since measurements are less complicated in ambient conditions, CaIrO$_3$ has been used extensively as
an analogue of MgSiO$_3$ post-perovskite for structural, mechanical, and transport studies 
\cite{pPv_cairo3_miyajima06,pPv_cairo3_martin07,pPv_cairo3_ballaran07,pPv_cairo3_hunt12}.
To unveil the valence electronic structure of CaIrO$_3$ we here carry out {\it ab initio} wave-function 
quantum chemical calculations.
With no SOC's, splittings as large as 0.8 eV are found for the $t_{2g}$ levels of CaIrO$_3$.
The $5d^5$ relativistic ground-state wave function therefore displays a highly uneven admixture of
$xy$, $xz$, and $yz$ character, which contradicts the interpretation of recent resonant x-ray
diffraction (RXD) experiments \cite{CaIrO3_ohgushi_2011}.
For the corner-sharing octahedra, the quantum chemical calculations further predict large AF
superexchange couplings \cite{superexchange_anderson59} of 120 meV, about the same size with the
magnetic interactions in the parent compounds of the high-$T_{\rm c}$ cuprate superconductors.
The lowest $d$-$d$ excited states are two doublets at approximately 0.7 and 1.4 eV, originating from the 
$j\!=\!3/2$ quartet in an ideal cubic environment.
This is the largest splitting reported so far for the $j\!=\!3/2$ quartet in $d^5$ systems
and singles out CaIrO$_3$ as an unique material in which both the SOC and the non-cubic crystal
field splittings are of the order of half eV or larger.


To investigate the local Ir $d$-level splittings, multiconfiguration self-consistent-field
(MCSCF) and multireference configuration-interaction (MRCI) calculations \cite{bookQC_2000}
are performed on embedded clusters made of one reference IrO$_6$ octahedron, ten adjacent Ca
sites, and four nearest-neighbor (NN) IrO$_6$ octahedra.
That a realistic representation of ions at NN polyhedra guarantees highly accurate results for
the $d$-level electronic structure at the central site was earlier pointed out for correlated $3d^1$, 
$3d^2$, and $3d^9$ oxide compounds \cite{CuO2_dd_hozoi11,TiVoxychlorides_dd_bogdanov11,QP_bands_cupr_hozoi07}.
The farther solid-state environment is modeled as a one-electron effective potential which in
an ionic picture reproduces the Madelung field in the cluster region.
All calculations are performed with the {\sc molpro} quantum chemical software \cite{molpro_brief}.
We employed energy-consistent relativistic pseudopotentials for Ir \cite{ECP_Stoll_5d} and Ca
\cite{ECP_Stoll_Ca_group} and Gaussian-type valence basis functions
\cite{ECP_Stoll_5d,ECP_Stoll_Ca_group,GBas_molpro_2p,ANOs_pierloot_95} (for details, see Supplementary Material \cite{ CaIrO3_supmat}).
For the ground-state calculations, the orbitals within each finite cluster are variationally 
optimized at the MCSCF level.
All Ir $t_{2g}$ functions are included in the active orbital space. 
Excitations within the Ir $t_{2g}$ shell are afterwards computed just for the central IrO$_6$
octahedron while the occupation of the NN Ir valence shells is held frozen as in the MCSCF
ground-state configuration.
Spin-orbit interactions are further accounted for as described in Ref.~\cite{SOC_molpro}.
The subsequent MRCI treatment includes all single and double excitations from the O $2p$ orbitals
at the central octahedron and the Ir $5d$ functions.
To partition the O $2p$ valence orbitals into two different groups, i.e., at sites of the 
central octahedron and at NN octahedra, we employ the orbital localization module available with
{\sc molpro}.
Crystallographic data as reported by Rodi and Babel \cite{CaIrO3_rodi_1965} is used.

\begin{table}[!t]
\caption{
Ir $t_{2g}$ splittings and relative energies of the spin-orbit $t_{2g}^5$ states in CaIrO$_3$.
MCSCF and MRCI results with and without SOC's for 5-plaquette clusters, see text.
}
\label{dd_CaIrO3}
\begin{ruledtabular}
\begin{tabular}{lcc}
$t_{2g}^5$ splittings (eV)
                      &MCSCF       &MRCI       \\
\hline
\\
$d_{xy}\!-\!d_{yz}$   &0.63        &0.68       \\
$d_{xy}\!-\!d_{xz}$   &0.76        &0.83       \\
\\
$|j\!=\!1/2, m_j\!=\!\pm\!1/2\rangle$
                      &0.00--0.12  &0.00--0.16 \\
$|j\!=\!3/2, m_j\!=\!\pm\!3/2\rangle$
                      &0.62--0.70  &0.66--0.76 \\
$|j\!=\!3/2, m_j\!=\!\pm\!1/2\rangle$
                      &1.25--1.27  &1.36--1.39 \\
\end{tabular}
\end{ruledtabular}
\end{table}

With no SOC's, we find Ir $d_{xy}\!-\!d_{yz}$ and $d_{xy}\!-\!d_{xz}$ splittings of 
0.63 and 0.76 eV by MCSCF calculations and slightly larger values of 0.68 and 0.83
eV at the MRCI level, with the $xy$ level the lowest (see the first lines in Table~\ref{dd_CaIrO3}).
The $j\!=\!1/2$ ground-state wave function therefore has dominant $xz$ character,
as much as $75\%$ by MRCI calculations with SOC (MRCI+SOC). 
The four components of the $j\!=\!3/2$ quartet are split into groups of two by 0.6--0.7 eV
(lower lines of Table~\ref{dd_CaIrO3}).
The lower two $j\!=\!3/2$ terms have $72\%$ $yz$ character, the highest $j\!=\!3/2$ components
have $90\%$ $xy$ character.

When SOC's are not accounted for, the coupling to spin moments at NN Ir sites gives rise for
each particular orbital occupation to one sextet, four quartet, and five doublet states. 
All those configuration state functions enter the spin-orbit calculations.
32 different spin-orbit states are therefore generated for each $j$ configuration at the 
central Ir site and this causes a finite energy spread in Table~\ref{dd_CaIrO3}. 
A simpler and more transparent picture can be obtained by replacing the four Ir$^{4+}$
$d^5$ NN's by closed-shell Pt$^{4+}$ $d^6$ ions. 
In that case, the MRCI excitation energies for the $j\!=\!3/2$ states at the central Ir
site are 0.67 and 1.38 eV.
An instructive computational experiment which we can further make is to replace the distorted
post-perovskite structure by a hypothetical cubic perovskite lattice in which the Ir-O bond
length is the average of the three different bond lengths in post-perovskite CaIrO$_3$, i.e.,
2.03 \AA .
This allows to directly extract the SOC $\lambda$ for the Ir$^{4+}$ $5d^5$ ion because the
splitting between the spin-orbit $j\!=\!1/2$ doublet and the four-fold degenerate $j\!=\!3/2$
state is $3\lambda/2$ in cubic symmetry \cite{book_abragam_bleaney}.
The doublet-quartet splitting comes out as 0.70 eV by MRCI+SOC calculations for the idealized
cubic perovskite structure.
This yields a spin-orbit coupling constant $\lambda\!=\!0.47$, in agreement with values of 0.39--0.49 eV earlier
extracted for Ir$^{4+}$ impurities from electron spin resonance measurements \cite{lambda_Ir_andlauer76}.

The results of our {\it ab initio} calculations, i.e., the unusually large $t_{2g}$  
splittings and the highly uneven admixture of $xz$, $yz$, and $xy$ character for the
spin-orbit $t_{2g}^5$ states, are in sharp contrast to the interpretation of recent
RXD experiments on CaIrO$_3$ \cite{CaIrO3_ohgushi_2011}.
In particular, the authors of Ref.~\cite{CaIrO3_ohgushi_2011} conclude from 
the RXD data that the deviations from an isotropic $xz$-$yz$-$xy$ picture are marginal,
as also found for instance in the layered square-lattice material Sr$_2$IrO$_4$ \cite{214Ir_kim_2009}.
While for Sr$_2$IrO$_4$ we also obtain \cite{214Ir_vmk_2012} rather small $t_{2g}$ splittings
of about 0.1 eV, comparable weights of the $xz$, $yz$, $xy$ components in the $j\!=\!1/2$ 
ground-state wave function, and NN superexchange interactions in good agreement with resonant
inelastic x-ray scattering (RIXS) data, for the post-perovskite CaIrO$_3$ the quantum 
chemical results point in a different direction.
These conflicting findings urgently call for Ir $L$-edge RIXS measurements on CaIrO$_3$.
The RIXS experiments have recently emerged as a most reliable technique for exploring the
charge, spin, and orbital degrees of freedom of correlated electrons in solids
\cite{RIXS_RevModPhys11,spin_orb_sep_schlappa12}.

\begin{figure}[b!]
\includegraphics[width=1\columnwidth]{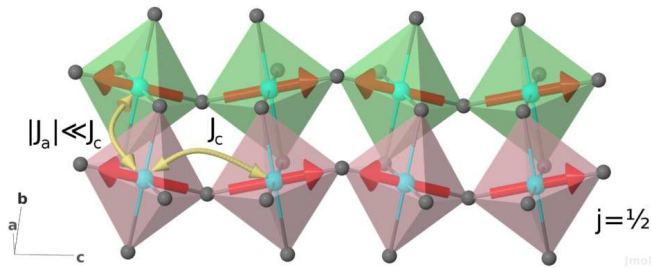}
\caption{Sketch of the NN magnetic interactions in CaIrO$_3$.
}
\label{fig02}
\end{figure}

\begin{table}[!t]
\caption{
NN effective magnetic couplings in CaIrO$_3$, see text.
Positive values denote AF exchange.
}
\label{J_CaIrO3}
\begin{ruledtabular}
\begin{tabular}{crrr}
$J$ (meV)          &MCSCF    &MRCI      &MRCI+SOC \\
\hline
\\
$J_{\rm c}$        &$91.2$   &$163.4$   &$121.0$  \\
$J_{\rm a}$        &$-2.3$   &$4.2$     &$-7.3$   \\
\end{tabular}
\end{ruledtabular}
\end{table}

The presence of both corner-sharing and edge-sharing octahedra and the large splittings within
the $t_{2g}$ shell have important implications on the nature and the magnitude of the magnetic
interactions.
To determine the latter, we designed 8-octahedra clusters including two active Ir$^{4+}$ $d^5$
sites.
The six $5d$ NN's were modeled as closed-shell Pt$^{4+}$ $d^6$ ions.
All possible occupations were allowed within the set of $t_{2g}$ orbitals at the two active
Ir sites in the MCSCF calculations, which gives rise to nine singlet and nine triplet states
with no SOC.
The MCSCF wave functions were expressed in terms of orbitals optimized for an average of
those singlet and triplet states.
All eighteen singlet and triplet states entered the spin-orbit calculations, both at the MCSCF and MRCI
levels.
In MRCI, single and double excitations from the Ir $t_{2g}$ orbitals and the $2p$ orbitals of 
the bridging ligand were included. Such a computational scheme provides $J$ values in very good
agreement with RIXS measurements in the layered iridate Sr$_2$IrO$_4$ \cite{214Ir_vmk_2012}.

Remarkably, with no SOC's, the NN AF exchange constant $J_{\rm c}$ for octahedra sharing
corners along the $c$ axis is as large as 91.2 meV by MCSCF calculations and 163.4 meV by
MRCI, see Table~\ref{J_CaIrO3}.
The $t_{2g}$ hole orbitals are here the $xz$ components, as discussed above.
Although the IrO$_6$ octahedra are tilted somewhat due to rotations about the $a$ axis, see
Fig.~1, this does not affect much the $\pi$-type $5d_{xz}$-$2p_x$-$5d_{xz}$ overlap and 
superexchange \cite{superexchange_anderson59} interactions.
An AF $J$ of 163.4 meV is even larger than the values for $\sigma$-type superexchange paths
in two-dimensional $S\!=\!1/2$ $3d$ Cu oxide superconductors \cite{CuO_J_Illas_00} and points
to the role of the larger spatial extent of $5d$ functions in iridates.
With SOC's accounted for, the effective AF coupling $J_{\rm c}$ is reduced to 121.0 meV 
(last column in Table~\ref{J_CaIrO3}). 
The spin-orbit ground-state wave function now acquires sizable $yz$ character and since
the tilting of the IrO$_6$ octahedra does affect the out-of-plane $5d_{yz}$-$2p_y$-$5d_{yz}$
superexchange path, the effective magnetic coupling between the $j\!=\!1/2$ sites is
smaller.

For edge-sharing octahedra along the $a$ axis, we find a weak NN ferromagnetic (FM) coupling
$J_{\rm a}\!\approx -7.3$ meV in the spin-orbit MRCI calculations.
The three components of the $j_{\rm tot}\!=\!1$ triplet state in the two-site problem display
small splittings of $\approx\!1$ meV (see Supplementary Material \cite{ CaIrO3_supmat}), which shows that anisotropic 
non-Heisenberg terms \cite{kitaev_213Ir_chaloupka10} should also be considered when mapping the 
{\it ab initio} data onto an effective spin Hamiltonian.
In estimating the Heisenberg coupling constant $J_{\rm a}$ we neglected such terms and 
used an average of the energies of the $j_{\rm tot}\!=\!1$ triplet components,
$J\!\approx\!\tilde E(j_{\rm tot}\!=\!1)\!-\!E(j_{\rm tot}\!=\!0)$.
The {\it ab initio} results, i.e., strong AF interactions along $c$ and weak FM couplings 
along $a$ (see Fig.~\ref{fig02}), are in full agreement with the observed striped AF magnetic ordering and very
large Curie-Weiss temperature \cite{CaIrO3_ohgushi_2011}.
The MRCI+SOC $J$ values thus provide a firm quantitative basis in understanding the magnetic
properties of this material.

We have employed, in summary, a set of many-body quantum chemical calculations to unravel the $5d$ electronic
structure of the post-perovskite iridate CaIrO$_3$. 
While electronic structure calculations have been earlier performed within the local density 
approximation to density functional theory and the role of electron correlation effects
has been pointed out \cite{CaIrO3_subedi_2012}, we here make clear quantitative predictions for
the Ir $t_{2g}$ splittings, character of the entangled spin-orbit wave function, and
magnitude of the NN superexchange interactions.
Our results single out CaIrO$_3$ as a $5d^5$ system in which lattice distortions and
spin-orbit couplings compete on the same energy scale to give rise to highly anisotropic
electronic structure and magnetic correlations.
In particular, the present {\it ab initio} investigation yields a different picture than previously derived
on the basis of resonant x-ray diffraction experiments for the relativistic ground-state wave
function.
The large $t_{2g}$ splittings that we have found from the calculations give rise to dominant $xz$ hole character and 
remarkably strong AF interactions along the $c$ axis.
Further, the large AF couplings along $c$ and weak FM exchange along $a$ characterize CaIrO$_3$ as a
$j\!=\!1/2$ quasi-one-dimensional antiferromagnet.

\begin{acknowledgments}
This paper is dedicated to Prof. Ria Broer to honour her devoted career in Theoretical and Computational Chemistry. 
N.~A.~B. and L.~H. acknowledge financial support from the Erasmus Mundus Programme of the European Union 
and the German Research Foundation (Deutsche Forschungsgemeinschaft, DFG), respectively.
\end{acknowledgments}

\newpage

\section{Supplementary Material}

\section{Computational details}

\subsection{Ir $d$-level splittings}
To determine the splittings within the Ir $t_{2g}$ levels and spin-orbit $t_{2g}^{5}$ states we employed 
large 5-octahedra clusters embedded in arrays of point charges fitted to reproduce the crystal Madelung field in
the cluster region.
Quadruple-zeta basis sets from the {\sc molpro} library were applied for the valence shells of the central
Ir ion \cite{SM_ECP_Stoll_5d} and triple-zeta basis sets for the ligands \cite{SM_GBas_molpro_2p} of the central octahedron 
and the nearest-neigbor Ir sites \cite{SM_ECP_Stoll_5d}.
For the central Ir ion we also used two polarization $f$ functions \cite{SM_ECP_Stoll_5d}.
For farther ligands in our clusters we applied minimal atomic-natural-orbital basis sets
\cite{SM_ANOs_pierloot_95}.
All occupied shells of the Ca$^{2+}$ ions were incorporated in the large-core pseudopotentials
and each of the Ca $4s$ orbitals was described by a single contracted Gaussian function \cite{SM_ECP_Stoll_Ca_group}.

\begin{table}[h]
\caption{
Relative energies (meV) of singlet and triplet states for two adjacent Ir sites for
corner-sharing octahedra. 
The singlet state is taken in each case as reference.
}
\label{J_c}
\begin{ruledtabular}
\begin{tabular}{crrr}
                   &MCSCF    &MRCI      &MRCI+SOC \\
\hline
\\
Singlet            &$0.0$    &$0.0$     &$0.0$    \\
Triplet            &$91.2$   &$163.4$   &$120.0$  \\
                   &         &          &$120.2$  \\
                   &         &          &$122.8$  \\
\end{tabular}
\end{ruledtabular}
\end{table}

\begin{table}[h]
\caption{
Relative energies (meV) of singlet and triplet states for two adjacent Ir sites for
edge-sharing octahedra.
The triplet or lowest triplet component is taken in each case as reference.
}
\label{J_a}
\begin{ruledtabular}
\begin{tabular}{crrr}
                   &MCSCF  &MRCI      &MRCI+SOC \\
\hline
\\
Triplet            &$0.0$  &$0.0$     &$0.0$    \\
                   &       &          &$0.4$    \\
                   &       &          &$1.5$    \\
Singlet            &$2.3$  &$-4.2$    &$7.9$    \\
\end{tabular}
\end{ruledtabular}
\end{table}

\subsection{Effective magnetic couplings}
Calculations for the isotropic magnetic exchange interactions were performed on 8-octahedra clusters
with two active Ir$^{4+}$ $5d^5$ sites.
To make the analysis of the intersite spin couplings straightforward, the six Ir$^{4+}$ $t_{2g}^{5}$ NN's were
replaced by closed-shell Pt$^{4+}$ $t_{2g}^{6}$ species.
The basis functions used here were as described above.
The only exception was the O ligand bridging the two magnetically active Ir sites, for which we employed
quintuple-zeta valence basis sets and four polarization $d$ functions \cite{SM_GBas_molpro_2p}.
Relative energies for the lowest $S_{\rm tot}\!=\!0$ and $j_{\rm tot}\!=\!0$ singlet and
$S_{\rm tot}\!=\!1$ and $j_{\rm tot}\!=\!1$ triplet states for corner-sharing and edge-sharing octahedra
are listed in Table~\ref{J_c} and Table~\ref{J_a}, respectively.
In the spin-orbit calculations the triplet components display small splittings that indicate the presence
of small non-Heisenberg terms.
In deriving the effective Heisenberg coupling constants $J_{\rm c}$ and $J_{\rm a}$ we neglected such terms and
used an average of the energies of the $j_{\rm tot}\!=\!1$ triplet components,
$J\!\approx\!\tilde E(j_{\rm tot}\!=\!1)\!-\!E(j_{\rm tot}\!=\!0)$.



\onecolumngrid

%

\end{document}